\documentclass[aps,pre,twocolumn,showpacs,floatfix,superscriptaddress, graphics, longbibliography]{revtex4-1}

\usepackage{amsmath,amssymb,graphicx,color,braket,hyphenat,makeidx,xcolor,dsfont,subfigure}
\usepackage[ocgcolorlinks,colorlinks=true,linkcolor=blue,citecolor=red,linktocpage=true]{hyperref}

\newcommand{\blue}{\color{\blue}}

\newcommand{\tr}{\mathrm{Tr}}

\begin{document}

\title{Optimal work extraction and thermodynamics of quantum measurements and correlations}

\author{Gonzalo Manzano}
\affiliation{Institute for Cross-Disciplinary Physics and Complex Systems IFISC (CSIC-UIB), Campus Universitat Illes Balears, E-07122 Palma de Mallorca, Spain}
\affiliation{Scuola Normale Superiore, Piazza dei Cavalieri 7, I-56126, Pisa, Italy}
\affiliation{International Center for Theoretical Physics ICTP, Strada Costiera 11, I-34151, Trieste, Italy}

\author{Francesco Plastina}
\affiliation{Dip.  Fisica, Universit\`a della Calabria, 87036 Arcavacata di Rende (CS), Italy}
\affiliation{INFN - Gruppo collegato di Cosenza, Cosenza Italy}

\author{Roberta Zambrini}
\affiliation{Institute for Cross-Disciplinary Physics and Complex Systems IFISC (CSIC-UIB),
Campus Universitat Illes Balears, E-07122 Palma de Mallorca, Spain}

\begin{abstract}
We analyze the role of indirect quantum measurements in work extraction from quantum systems in nonequilibrium states.
In particular, we focus on the work that can be obtained by exploiting the correlations shared between the system of interest and an additional ancilla,
where measurement backaction introduces a nontrivial thermodynamic tradeoff. We present optimal state-dependent protocols for extracting work from both
classical and quantum correlations, the latter being measured by discord. Our quantitative analysis establishes that, while the work content of classical correlations can be fully
extracted by performing local operations on the system of interest, accessing work related to quantum discord requires a specific driving protocol
that includes interaction between system and ancilla.
\end{abstract}

 \maketitle 

\section{Introduction}

One of the aims of quantum thermodynamics
\cite{john,janet} is the precise identification of the role of genuine
quantum resources, such as coherence \cite{cohe}, correlations \cite{corre} or squeezing \cite{squeez}, both in the
performance of thermodynamic tasks by nano-machines \cite{quama1,quama2,quama3}, and, on a more fundamental ground, in the description of finite-time
non-equilibrium thermodynamic processes \cite{noi,corre2,noneq}. In this context, interest has been raised toward the process of work extraction
from quantum systems \cite{All04,Skr14}, and on its enhancement in feedback protocols \cite{extradem}.
Although entanglement generation is not essential for optimal work extraction  from quantum batteries \cite{Generation},
it can, nevertheless, be exploited to increase the amount of extractable work \cite{Funo13, alicki, Acin15}. Quantum discord  can enter prominently
in enhancing the performance of Maxwell demons \cite{Demons}, heat engines \cite{Machines} and work extraction protocols \cite{glsteve} as well,
and both entanglement and discord have been shown to play a role in the work gain obtained thanks to a feedback enhanced extraction protocol \cite{Fra17}.
However, obtaining quantitative connections between the extracted work and both classical and quantum correlations has been shown very challenging,
since these may depend on the allowed operations \cite{Acin15}.

The prototypical scenario considers a cyclic unitary transformation to extract work from a quantum system $S$ \cite{All04}. This case has been extended
by considering that $S$ shares correlations with an ancilla $A$ \cite{Fra17}. The unitary transformation setting has been generalized as well,
by including thermalization processes,\cite{Esp11, Par15}. The scope of this letter is to obtain quantitative relations between the optimal work gain and classical
and quantum correlations in this general framework of work extraction, where both access to a thermal reservoir is allowed \cite{Esp11, Par15}, and feedback is provided
by a measurement performed on $A$.
We will show that a tight link exists between the optimal work gain obtainable in presence of feedback and the {\it classical} correlations.
Turning to the role of quantum correlations, we will introduce a work contribution due to quantum discord, arguing that it cannot
be extracted in a feedback enforced protocol, as it is unavoidably lost after a local measurement is performed.
However, an improved protocol can be designed, in which the work content of quantum discord can be possibly extracted {\it before} the measurement and
the feedback enhanced protocol are performed.
In doing this, we will elucidate the role and the energetic value of both classical and quantum correlations and, at the same time,
discuss the energetic cost of the measurement that is necessary to provide feedback. Contrary to the classical case,
quantum measurements introduce a tradeoff between the gain in extractable
work due to the measurement-induced local entropy reduction, and its loss
due to correlations erasure. Despite this, we will show that the total amount
of work extractable with the generalized protocol can overcome
the one obtained without measurement and feedback, provided optimal measurements are performed.

\section{Setup}

We consider work extraction from a quantum system
with Hamiltonian $H_S$ in an arbitrary nonequilibrium state
$\rho_S$. In the most simple situation, only unitary operations
are allowed, where the Hamiltonian changes according to some
cyclic protocol, in which $H_S$ is the same Hamiltonian before and
after the operation. In such a case, the maximum work that can be
extracted from an initial (non-passive) state
$\rho_S$ is the so-called ergotropy $\mathcal{W}_S$ \cite{All04}.
This framework can be naturally extended, with a performance enhancement, by
including also non-unitary transformations. In
particular, if, in addition to unitary cyclic driving,
contact with a thermal reservoir is also allowed, the extracted work may increase due to both the system
entropy varying during the protocol, and to the reservoir providing some extra energy.
In this case, the maximum
amount of extractable work is given by the difference in
nonequilibrium free energy between the state $\rho_S$ and the
thermal equilibrium state at the reservoir inverse temperature
$\beta \equiv 1/k_B T$ \cite{Esp11, Par15}
\begin{equation} \label{eq:maxwork}
W_\mathrm{ext} \leq \mathcal{W}^{\beta}_S =
\mathcal{F}_{\beta}(\rho_S) - \mathcal{F}_{\beta}(\rho_S^{\beta})
= k_B T D(\rho_S || \rho_S^{\beta}),
\end{equation}
where $\rho_S^{\beta}=e^{- \beta H_S}/Z_S$ is the equilibrium
state, while $D(\rho || \sigma) = \tr[\rho (\log \rho - \log
\sigma)]$ stands for the quantum relative entropy \cite{Sag13,Ved02}.
The nonequilibrium free energy for a system in state $\rho_S$,
with Hamiltonian $H_S$, and with respect to a thermal bath at
temperature $T$ is defined as
\begin{equation}
\mathcal{F}_{\beta}(\rho_S) = \tr[ H_S \rho_S] - k_B T S(\rho_S),
\end{equation}
where $S(\rho)= - \tr[\rho \ln \rho]$ is the von Neumann entropy,
and where, for thermal states $\mathcal{F}_{\beta}(\rho_S^{\beta})
= - k_B T \ln Z_S$ reduces to the Helmholtz free energy.
Equality in Eq. \eqref{eq:maxwork} may be obtained by implementing
an operationally reversible isothermal process \cite{Skr14, Esp11, Par15}.
This is made up of two steps: first, a sudden quench is
performed, in which the Hamiltonian $H_S$ is changed into
$H_{\rho_S} = -k_B T \ln \rho_S$; then, a quasi-static isothermal
transformation follows, during which the Hamiltonian turns back to
$H_S$, while the system is kept in contact with the heat bath. 
In this second step, the system always stays in equilibrium with the
reservoir, ending up in the state $\rho_S^{\beta}$ \cite{Esp11,Par15}.
Notice that here it is assumed that the thermal reservoir always induces decoherence and dissipation in the instantaneous energy eigenbasis \cite{noteref}. 
Such an isothermal transformation can be constructed by means of an infinite sequence of quantum maps acting over
infinitesimal time-steps (the demostration is left to \cite{supple}). This optimal isothermal work extraction procedure always
outperforms cyclic unitary protocols: independently of the temperature, one can show that the decrease in free energy
is larger than the ergotropy, $\mathcal{W}_S^{\beta} \geq \mathcal{W}_S$, $\forall \beta$, where the equality is achieved only when
the temperature verifies $S(\rho_S^\beta) = S(\rho_S)$ (the proof is given in \cite{supple}). Notice that the presence of the environment
plays here a constructive role, allowing an extra source of energy and increasing our ability to extract work.

In the following, we will extend the optimal isothermal protocol to
the case in which the system of interest $S$ 
is prepared in a joint state $\rho_{SA}$ with an uncoupled
ancillary system $A$
, with which it may share classical and/or quantum correlations.
Specifically, the total amount of correlations between the two
parts can be measured by the quantum mutual information
$I(\rho_{SA}) = D(\rho_{SA} || \rho_S \otimes \rho_A) \geq 0$,
where $\rho_S = \tr_A[\rho_{SA}]$ and $\rho_A = \tr_S[\rho_{SA}]$
are the marginal (reduced) states.

We will first show that the amount of work extractable from the
system of interest increases when some information is
provided after a measurement is performed on the ancilla. In this
way, we provide both a generalization of the nonequilibrium isothermal work extraction setup to include
quantum measurement induced feedback, and a generalization to the case of
entropy-changing transformations of the result of
Ref. \cite{Fra17} for the ergotropy. 

\begin{figure}[t]
\includegraphics[width= 0.40 \textwidth]{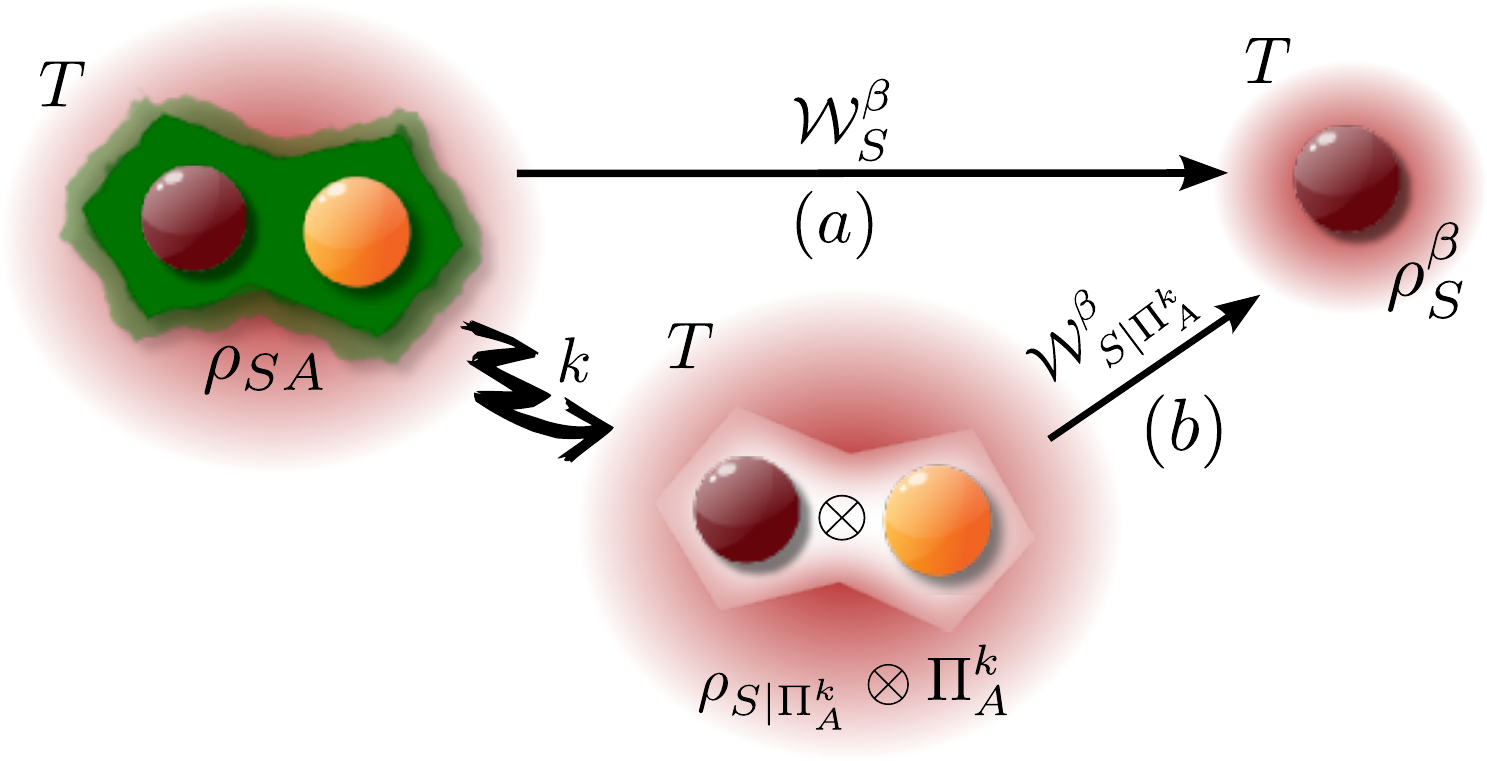}
\centering
\caption{(color online) Starting from $\rho_{SA}$, work can be extracted from $S$ either by a direct isothermal protocol [path (a)], or by first
performing a measurement on $A$, and then applying an outcome dependent isothermal protocol [path (b)].} \label{figure1}
\end{figure}

To start with, let us define $\mathcal{W}^{\beta}_{S|\pi_A}$ as
the maximum amount of work extractable from $S$ by exploiting the feedback obtained from a measurement
performed on $A$. In particular, we consider a projective
measurement, described by the set of projectors
$\pi_A=\{\Pi_A^k \}$ for $k= 1,..., d_A$. After optimizing the extracted work
over all possible sets of projectors $\pi_A$ (that is, over all
possible measurements on $A$), we define
$\mathcal{W}^{\beta}_{S|A} = \mbox{max}_{\pi_A}
\mathcal{W}^{\beta}_{S|\pi_A}$ (see Fig. \ref{figure1}).
Therefore, during this first part, we take as an operational assumption that work is only extracted from the system and not from the ancilla.

The measurement affects both the ancilla and the system state. In particular, if the outcome $k$ occurs
(with probability $p_k =\tr[\Pi_A^k \rho_{S A}]$), then the state of the system is updated to
\begin{equation} \label{eq:backaction} \rho_{S} ~~\longrightarrow
~~\rho_{S|\Pi_A^k} = \tr_A[(\mathbb{I}_S \otimes \Pi_A^k) \rho_{SA} (\mathbb{I}_S \otimes \Pi_A^k)]/p_k,
\end{equation}
$\mathbb{I}_S$ being the identity operator for $S$. Notice that the measurement operators commute with the Hamiltonian of the system, $[H_S, \mathbb{I}_S \otimes \Pi_{A}^k] = 0 ~\forall k$;
therefore the system energy remains constant during the measurement process.


For an initially uncorrelated $SA$ state, that
is, for $\rho_{SA} = \rho_S \otimes \rho_A$ --- or equivalently
$I(\rho_{SA}) = 0$ --- measurements on the ancilla do not induce any
change in the system state.

\section{Extracting work from classical correlations}

In the
optimal isothermal protocol discussed above, the maximum work that
can be extracted on average from $S$ is increased at best by the
amount of {\it classical} correlations initially present in the state
$\rho_{SA}$.

To show this, we start by considering that, if the ancilla is
subjected to the projective measurement introduced above, when
outcome $k$ is obtained, $S$ suffers from the back-action
corresponding to Eq. \eqref{eq:backaction}, and this amounts to a
change in the system's free energy
\begin{equation} \label{eq:freechange}
 \Delta \mathcal{F}^{\beta}_{S| \Pi_A^k} = \mathcal{F}_{\beta}(\rho_{S|\Pi_A^k}) - \mathcal{F}_{\beta}(\rho_{S}).
\end{equation}
The outcome $k$ being known, we may adapt the optimal isothermal
protocol introduced above, which now will depend on $k$:
in the first step, a $k$-dependent sudden quench of the
Hamiltonian is performed, with $H_S \rightarrow H_{\rho_{S| \Pi_A^k}} =
-k_B T \ln \rho_{S|\Pi_A^k}$, which is then quasi-statically brought
back to $H_S$ while in contact with the thermal reservoir. This
process requires precise knowledge of the state $\rho_{S|\Pi_A^k}$,
which in turn implies knowing the initial state $\rho_{SA}$ and
the set of projectors $\pi_A$.

With the same argument recalled above, one may conclude that the
maximum amount of work extractable from the state
$\rho_{S|\Pi_A^k}$ is given by ${\cal W}^{\beta}_{S|\Pi_A^k} \equiv
\mathcal{F}_{\beta}(\rho_{S|\Pi_A^k}) -
\mathcal{F}_{\beta}(\rho_S^{\beta})$. The average work extracted
after many repetitions of this process is, then,
\begin{align} \label{eq:maxwork2}
\mathcal{W}^{\beta}_{S|\pi_A} =  \sum_k p_k
\mathcal{W}^{\beta}_{S|\Pi_A^k} = \sum_k p_k \Delta
\mathcal{F}^{\beta}_{S|\Pi_A^k} + \mathcal{W}^{\beta}_S,
\end{align}
where, in the last equality, we used Eqs. \eqref{eq:freechange}
and \eqref{eq:maxwork}. A sketch of the protocol and of this result is given in Fig. \ref{figure1}.

The average change of the generalized free energy, entering the Eq. \eqref{eq:maxwork2} above, has a clear information theoretic interpretation when it is
expressed in terms  of the entropy change: $\sum_k p_k \Delta
\mathcal{F}^{\beta}_{S|\Pi_A^k} =  k_B T [ S(\rho_S) - \sum_k p_k S(\rho_{S|\Pi_A^k}) ] \equiv k_B T J(\rho_{SA})_{\pi_A}$.
The quantity $J(\rho_{SA})_{\pi_A}$ gives the mutual information extracted by the local measurement performed on $A$ by using the
set of projectors $\pi_A=\{ \Pi_A^k \}$ \cite{Zur01, Hen01}. The same quantity has been employed to discuss feedback controlled protocols in Ref. \cite{Sag08}.

As a result, the average increase of the work extracted during the process reads
\begin{equation} \label{eq:maxwork3}
\Delta \mathcal{W}^{\beta}_{S| \pi_A} = \sum_k p_k \Delta \mathcal{F}^{\beta}_{S|\Pi_A^k}
= k_B T J(\rho_{SA})_{\pi_A} \geq 0,
\end{equation}
where $\Delta {\cal W}^{\beta}_{S| \pi_A}= {\cal W}^{\beta}_{S|\pi_A} - {\cal W}^{\beta}_{S}$ is the gain in extractable work, and where
the inequality in \eqref{eq:maxwork3} follows directly from
concavity of von Neumann entropy and implies that
an average enhancement in the extracted work is found for any measurement.
No gain is obtained only if $\rho_{SA}$ is factorized; while, if $S$ and $A$ are correlated to some extent,
the extractable work can increase thanks to the feedback coming from the knowledge of the measurement outcome.
Intuitively, this is due to the fact that a measurement can increase the free energy of $S$ \cite{Par15}.

When optimized over all possible measurements, the quantity $J(\rho_{SA})_{\pi_A}$ introduced above gives a measure of the classical correlations
shared by $S$ and $A$ in the state $\rho_{SA}$, as defined in Refs. \cite{Zur01, Hen01}. There, a measurement oriented framework is put forward, and
classical correlations are defined as $J(\rho_{SA}) = \max_{\pi_A} J(\rho_{SA})_{\pi_A}$.

Thus, if we maximize Eq. \eqref{eq:maxwork3} over all sets of
projectors $\pi_A$, we obtain that the maximum enhancement in work
extraction $\Delta \mathcal{W}^{\beta}_{S|A} \equiv \max_{\pi_A}
\Delta \mathcal{W}^{\beta}_{S| \pi_A}$ is precisely given by
\begin{equation} \label{eq:result1}
\Delta \mathcal{W}^{\beta}_{S|A}  = k_B T J(\rho_{SA}) ,
 \end{equation}
Eq. \eqref{eq:result1} is the first of our main results; it tells us that the
gain in the work extracted from $S$ thanks to the
feedback protocol in which $A$ is measured, is due to (and upper
bounded by) the classical correlations shared by $S$ and $A$.

Even if quantum correlations do not contribute to Eq.
\eqref{eq:result1}, this does not imply that they do not play any
role, as we will see in the remainder of this letter.

In a classical context, where measurement back-action can be avoided in principle, the correlation function $J$ would coincide with the mutual information $I$, stemming from Bayes' rule. In the quantum regime, the difference between these two classically equivalent definitions of mutual information, called discord \cite{Zur01, Hen01}, gives a measure of the amount of non-classical correlations in the state $\rho_{SA}$ \cite{erre1}
\begin{equation} \label{eq:discord}  \mathcal{D}(\rho_{SA}) \equiv I(\rho_{SA}) - J(\rho_{SA}) \geq 0.
\end{equation}
Our result in Eq. (\ref{eq:result1}) gets a clear physical interpretation if discord is
used to understand it. From its definition, we can understand quantum discord as the
amount of correlations present in a bipartite quantum state, which
cannot be accessed by local measurements on one party.
Therefore, intuition dictates that as long as this information is
not available from measuring the ancilla $A$, it cannot  be used
in any way to improve our ability of extracting work from $S$.
More precisely, the work extractable from the whole $SA$ system
in the state $\rho_{SA}$ is given by the free-energy difference
between this state and the thermal reference one,
\begin{align} \label{eq:Wsa}
\mathcal{W}^{\beta}_{SA}(\rho_{SA}) &\equiv \mathcal{F}_{\beta}(\rho_{SA}) - \mathcal{F}_{\beta}(\rho_S^{\beta} \otimes \rho_A^{\beta}) \nonumber \\
       &= \mathcal{W}^{\beta}_S(\rho_{S}) + \mathcal{W}_A^{\beta}(\rho_{A}) + k_B T I(\rho_{SA}),
\end{align}
where $\mathcal{W}^{\beta}_A(\rho_{A}) =
\mathcal{F}_{\beta}(\rho_A) - \mathcal{F}_{\beta}(\rho_A^{\beta})
\geq 0$ is the work locally extractable from the ancilla $A$
without using measurements. From Eq. \eqref{eq:result1}, it then
follows that the work extractable from $S$ through the optimal
isothermal protocol supplemented by the feedback scheme,
$\mathcal{W}^{\beta}_{S|A}$, 
plus the work extractable from
$\rho_A$, can never exceed $\mathcal{W}^{\beta}_{SA}$:
\begin{align} \label{eq:discwork}
\mathcal{W}^{\beta}_{S|A} + \mathcal{W}^{\beta}_A(\rho_{A}) =
\mathcal{W}^{\beta}_{SA}(\rho_{SA}) - k_B T
\mathcal{D}(\rho_{SA}) .
\end{align}
Equation \eqref{eq:discwork} has a clear interpretation. The
intrinsic irreversibility of the measurement process  destroys the
quantum correlations present in state $\rho_{SA}$, as measured by
quantum discord. As a consequence, the work extractable from
system and ancilla decreases by an amount $k_B T
\mathcal{D}(\rho_{SA})$, which corresponds to the work value of
quantum correlations in the state $\rho_{SA}$.
Result \eqref{eq:result1} is then an exact expression stressing the deep link between work and knowledge. 
This interpretation of the role of discord agrees with that provided in Refs. \cite{Demons},
when comparing local and global Maxwell demon-like configurations.

\section{Thermodynamic tradeoff of quantum measurement}

In the above
discussion, we summed up the two extractable works
obtained i) from $S$, with the optimal protocol including feedback,
and, separately, ii) from $A$.
Although providing a nice interpretation for the work content of
quantum discord, this does not properly take into account the
measurement back-action on $A$, as
$\mathcal{W}_{\beta}^A$ would be the work extractable from $A$
if no measurement had been performed. In fact, the projective
measurement, in the first stage of the feedback scheme, modifies the whole
$SA$-state. After the $k$-th outcome, one has
\begin{equation}
 \rho_{SA} ~\longrightarrow~ \rho_{S|\Pi_A^k} \otimes \Pi_A^k,
\end{equation}
Then, one may ask how the work extracted from $SA$ in presence of
the feedback gets modified and wether it can in fact surpass
$\mathcal{W}^{\beta}_{SA}$ in Eq. \eqref{eq:Wsa}. To answer this
question, we consider the gain in work extraction obtained
from the true post-measurement state, with respect to $\mathcal{W}^{\beta}_{SA}$,  i.e. $\Delta
\mathcal{W}^{\beta}_{SA|\pi_A} =  \sum_k p_k
\mathcal{F}_{\beta}(\rho_{S|\Pi_A^k} \otimes \Pi_A^k) -
\mathcal{F}_{\beta}(\rho_{SA})$.

To perform a proper energy balance in presence of the measurement
process, we should also consider its work cost. If $H_A$ is the
Hamiltonian of the ancilla, and if $\rho_{A|\pi_A} = \sum_k p_k \Pi_A^k$
is its unconditional, post-measurement state, then the cost
$\mathcal{C}(\pi_A) \equiv \tr[H_A (\rho_{A|\pi_A} - \rho_A)]$
corresponds to the work needed to perform the measurement $\pi_A$ \cite{newla}.
It vanishes as soon as measurements are performed in the energy
eigenbasis, $[\Pi_A^k, H_A] = 0$, or when energy-less ancillas are
considered ($H_A \propto \mathbb{I}_A$) \cite{elouard}. More importantly, if the optimal
set of projectors $\pi_A^{\mathrm{opt}}$ is taken, which maximizes the
extracted classical information in Eq. \eqref{eq:result1}, we
have
\begin{align} \label{eq:workenh}
 \Delta \mathcal{W}^{\beta}_{SA|\pi_A^{\mathrm{opt}}} - & \mathcal{C}(\pi_A^{\mathrm{opt}}) = k_B T [S(\rho_{SA}) - \sum_k p_k S(\rho_{S|\Pi_A^k})] \nonumber \\
 &= k_B T [S(\rho_A) - \mathcal{D}(\rho_{SA})] \geq 0,
\end{align}
where the final inequality in Eq. \eqref{eq:workenh} follows from
the fact that discord is always bounded from above by the entropy
of the measured system \cite{Li11}.

The above Eq. \eqref{eq:workenh} is the second of our main results. It
remarkably ensures that the amount of extractable work from system
and ancilla does not decrease when using optimal quantum measurements and
feedback in the work extraction process, even if the cost of the
measurement is properly accounted for and subtracted. The
interpretation of the two terms above becomes clear if one notices
that the measurement induced free energy change can be written $\Delta \mathcal{F}^{\beta}_A \equiv \sum_k p_k
\mathcal{F}_{\beta}(\Pi_A^k) - \mathcal{F}_{\beta}(\rho_A) =
\mathcal{C}(\pi_A)+ k_B T S(\rho_A)$.
Thus, even if the quantum measurement produces a decrease in the
extractable work of the composite system by an amount $k_B T
\mathcal{D}(\rho_{SA})$, corresponding to the loss of quantum
discord, this is always (over-)compensated by an increase, $\Delta
\mathcal{W}_A^\beta = \Delta \mathcal{F}^{\beta}_A$, of the work locally
extractable from the ancillary system after the measurement. Indeed, this provides both a compensation for the
measurement cost, as well as the extra work amount $k_B T S(\rho_A)$, exceeding the work value of discord.
It is worth noticing here that if the optimal set of projectors $\pi_A^\mathrm{opt}$ were not used, then
$\Delta \mathcal{W}^{\beta}_{SA|\pi_A^{\mathrm{opt}}} \geq \mathcal{C}(\pi_A^{\mathrm{opt}})$ cannot be ensured anymore, and the tradeoff between
the gain in extractable work due to the measurement, and its reduction due to correlation erasure may give a detrimental result, implying that
the direct work extraction from $\rho_{SA}$ (without using measurements) is the best option.

\section{Extracting work from quantum correlations}

\begin{figure}[t]
\centering
\includegraphics[width= 0.45 \textwidth]{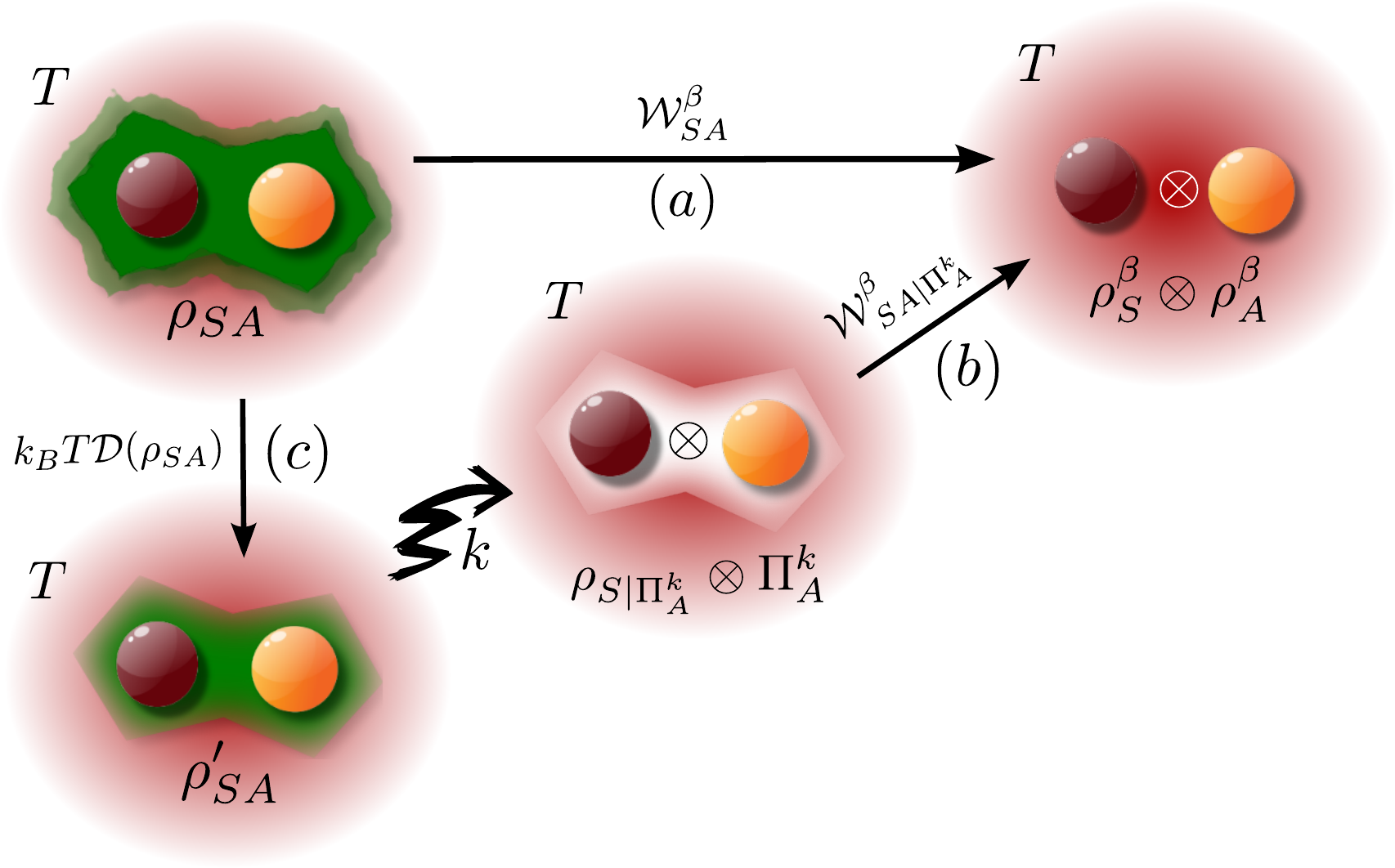}
\caption{(color online) The maximum amount of extractable work from $S$ and $A$ [path (a)] can be enhanced
using quantum measurements [path (b)]. However, the work content of quantum discord is in general lost in this case.
Optimal feedback enhanced work extraction can then be achieved by extracting it before the measurement is performed [step (c)].} \label{figure2}
\end{figure}

Finally, we are interested in the possibility of extracting the work content of quantum correlations,
that may also exceed classical correlations \cite{Gal11}, without renouncing to the benefits of the measurement.
This may seem impossible at first sight, as including projective quantum measurements will eventually produce the loss of discord in state $\rho_{SA}$, as we already discussed.
We propose a protocol for which this can be circumvented extracting the work content of quantum correlations {\it before} the projective measurement is performed on the ancilla.
This means including a new initial step $\rho_{SA} \rightarrow \rho_{SA}^\prime$ in the extraction protocol, performed before measurement and isothermal driving, sketched as step (c) in Fig. \ref{figure2}.
Such a step unavoidably requires interaction between system and ancilla.

Leaving considerations motivating the construction of this reversible sub-process to the Supplemental Material \cite{supple}, we require a final state of the
step with zero quantum correlations, but intact classical ones
\begin{equation}
 \rho_{SA}^\prime = \sum_k p_k \rho_{S|\Pi_A^k} \otimes \Pi_A^k.
\end{equation}
where, once again, the projectors $\Pi_A^k$ are taken from the optimal set $\pi_A^{\mathrm{opt}}$.
The step goes as follows: First we perform a sudden
quench of the total Hamiltonian, so that $H_S + H_A
\rightarrow H_{SA} \equiv -k_B T \ln \rho_{SA}$. Then, a
quasi-static driving is applied to the compound system,
transforming $H_{SA} \rightarrow H_{SA}^\prime \equiv - k_B T \ln
\rho_{SA}^\prime$, which leads the compound system to end up in
the state $\rho_{SA}^\prime$, as it follows from the fact that
$\rho_{SA}^\prime$ is now the equilibrium state at temperature $T$
with respect to the Hamiltonian $H_{SA}^\prime$. Finally, a second
sudden change of the Hamiltonian is performed $H_{SA}^\prime
\rightarrow H_{S} + H_{A}$. The maximum work extractable in this reversible
three-step process is, then,
\begin{equation} \label{eq:workenh2}
{\mathcal W}^{\beta}_{SA}(\rho_{SA}) -  {\mathcal
W}^{\beta}_{SA}(\rho_{SA}^\prime) = k_B T \mathcal{D}(\rho_{SA}).
\end{equation}


The full extraction of work is finally completed by applying the feedback enhanced protocol to $SA$ (see Fig. \ref{figure2}).
Summing up all of the contributions, the maximum work extractable from $\rho_{SA}$ is obtained by adding
the work value of discord [Eq. \eqref{eq:workenh2}], the one extractable directly from $\rho_{SA}^\prime$ [Eq. \eqref{eq:Wsa}],
plus the entropic gain due to the measurement [Eq. \eqref{eq:workenh} applied to $\rho_{SA}^\prime$]; that is,
\begin{align}\label{eq:finalwork}
 & {\mathcal W}_{SA|\pi_A^{\mathrm{opt}}}^{\beta}  (\rho_{SA}) -  \mathcal{C}(\pi_A^{\mathrm{opt}}) =  \\
 & \quad= k_B T {\mathcal D}(\rho_{SA}) + {\mathcal W}^{\beta}_{SA}(\rho_{SA}^\prime) +  k_B T S(\rho_A)  \nonumber \\
& \quad =  \mathcal{W}^{\beta}_S(\rho_{S}) + \mathcal{W}_A^{\beta}(\rho_{A}) + k_B T I(\rho_{SA}) + k_B T S(\rho_A). \nonumber
\end{align}
In particular, this implies that the process involving feedback from $A$ helps in increasing the extractable work even in comparison with the
optimal isothermal protocol (without feedback) applied to the whole $SA$ system.
This means that using a local quantum measurement may allow
not only to extract the full work associated to the total amount of
correlations present in $\rho_{AB}$, namely $k_B T I(\rho_{AB})$, but also an
enhancement proportional to the entropy of the ancilla. The latter, eventually, may
be lost in restoring the initial state of $A$ \cite{Sag08}. Finally, the recognition of the maximum extractable work in Eq. \eqref{eq:finalwork} allows for the
definition of a suitable information-to-work conversion efficiency in line with Ref. \cite{Efficiency}.

\section{Conclusions}

In conclusion, we derived quantitative relations linking the optimal work extractable from bipartite quantum systems and their classical and quantum
correlations, assessing both the role of thermal environments and quantum measurements. Moreover, we proposed a protocol to extract the work associated
to the presence of not only classical but also quantum correlations. Our results, beyond establishing a way to exploit quantum correlations thermodynamically,
might be of interest in practical applications regarding the design of quantum batteries \cite{Campisi}.

\begin{acknowledgments}
{\itshape Acknowledgements-}  We thank useful comments from Juan M. R. Parrondo. The authors acknowledge support from
the Horizon 2020 EU collaborative project QuProCS (Grant Agreement No. 641277). G. M. and R. Z. acknowledge MINECO/AEI/FEDER
through projects NoMaQ FIS2014-60343-P and EPheQuCS FIS2016-78010-P.
\end{acknowledgments}

\appendix

\section{Quasi-static isothermal processes with CPTP maps}
Here we explicitly construct a concatenation of CPTP maps leading to a generic isothermal quasi-static process as needed for reversible work extraction in the protocols
introduced in the main text. As mentioned there, an isothermal reversible process for a system with Hamiltonian $H$ and density operator $\rho_0$ can be constructed by 
a sudden quench, changing the Hamiltonian $H \rightarrow H_0 \equiv -k_B T \ln \rho_0$ (while leaving the system in state $\rho_0$), $T$ being the temperature of the bath, 
followed by a quasi-static process where $H_0$ is changed back to $H$ and the system remains in equilibrium with the reservoir at any time, ending thus in $\rho_S^\beta = e^{- \beta H}/Z$.
We notice that, because of the logarithm, the Hamiltonian $H_0$ as introduced before is only well defined for positive-definite initial states $\rho_0$. Nevertheless, initial states without full 
rank may be incorporated without spoiling our results. This can be done by simply including a slight modification in the definition of $H_0$, as we detail at the end of this section of the 
Supplemental Material. Therefore what follows apply to arbitrary initial states $\rho_0$.

In order to build the map describing the quasi-static process, we take inspiration from Ref. \cite{Isothermal} (see also Refs. \cite{Giovannetti,Uzdin} for similar analysis), where isothermal processes are constructed by means of alternating infinitesimal 
adiabatic and isochoric processes in an infinite sequence. Similarly, we assume, here, an infinite sequence of maps $\mathcal{E}_1 \circ \mathcal{E}_2 \circ \cdots \circ \mathcal{E}_N$, with 
$N \rightarrow \infty$, each of them describing an infinitessimal time step of the dynamics, with $\mathcal{E}_n(\rho_{n-1}) = \rho_n$, and the Hamiltonian changing as $H_{n-1} \rightarrow H_n$, 
where we set $H_N \equiv H$.

Let us decompose every CPTP map of the sequence in two steps
\begin{equation} \label{eq:maps}
\mathcal{E}_n (\rho) \equiv \mathcal{G}_n \circ \mathcal{U}_n (\rho),
\end{equation}
where $\mathcal{U}_n(\rho_{n-1})= \rho_{n-1}$ is a unitary sudden quench of the system Hamiltonian, $H_{n-1} \rightarrow H_n$, performed by the driving agent,
followed by a instantaneous Gibbs-preserving map verifiying $\mathcal{G}_n(\frac{e^{- \beta H_n}}{Z_n}) = \frac{e^{- \beta H_n}}{Z_n}$, which describes
the interaction with the environment. The Hamiltonian remains constant during this second step.

In order to ensure an isothermal reversible process, we need, for each CPTP map, that the change in the entropy of the system equals (minus) the heat introduced by
the environment divided by $k_B T$, that is
\begin{align} \label{eq:reversible}
\Delta S_n &\equiv S(\rho_n) - S(\rho_{n-1}) \nonumber \\ 
&= \beta \tr[H_n (\rho_n - \rho_{n-1})] \equiv - \beta Q_n,
\end{align}
where only $H_n$ appears in the expression above as the system only exchanges energy with the reservoir during the second step, $\mathcal{G}_n(\rho_{n-1}) = \rho_n$. This would require that the system state is always close to the
instantaneous equilibrium state for every step, $\frac{e^{-\beta H_n}}{Z_n}$. In the following we show that when assuming an infinitesimal change in the drive during
any step
\begin{equation} \label{eq:hamil}
 H_n = H_{n-1} + \epsilon \Delta H_n
\end{equation}
with $\epsilon \ll 1$, then the sequence of CPTP maps defined by Eq. \eqref{eq:maps}, verify the condition for reversibility, Eq. \eqref{eq:reversible}, up to first order in $\epsilon$, i.e.
irreversibilities come only to order $O(\epsilon^2)$.

We prove the above statement in two steps. First, we will show that if the state of the systems starts close to the equilibrium state before the map, then it remains close to the new equilibrium
state after the application of the map. That is
\begin{align} \label{eq:quasistatic}
 \rho_n &= \mathcal{G}_n\left(\frac{e^{- \beta H_{n-1}}}{Z_{n-1}} + \epsilon \Delta \rho_{n-1}\right) \nonumber \\ 
 &= \frac{e^{-\beta H_n}}{Z_n} + \epsilon \Delta \rho_n + O(\epsilon^2),
\end{align}
where $\Delta \rho_{n-1}$ and $\Delta \rho_{n}$ are tracelees operators. This ensures self-consistency of our construction. As a second step, we will prove that, since we may always rewrite $\rho_n = \rho_{n-1} + \epsilon \sigma_n$
for a suitable traceless $\sigma_n$ (in general $\Delta \rho_n \neq \sigma_n$), this implies Eq. \eqref{eq:reversible}.

We start introducing Eq.  \eqref{eq:hamil} into the left-hand-side of Eq. \eqref{eq:quasistatic}, which by using linearity and expanding $e^{\epsilon \beta \Delta H_n} = 1 + \epsilon \beta H_n + O(\epsilon^2)$,
and $Z_{n-1} = Z_n [1 + \epsilon \beta \tr[\Delta H_n] + O(\epsilon^2)]$ gives
\begin{widetext}
\begin{align}
& \mathcal{G}_n \left(\frac{e^{- \beta H_{n-1}}}{Z_{n-1}} + \epsilon \Delta \rho_{n-1} \right) = \mathcal{G}_n \left( \frac{e^{-\beta H_n}}{Z_n} \left[ \frac{1 + \epsilon \beta \Delta H_n + O(\epsilon^2)}{1 + \epsilon \beta \tr[\Delta H_n] + O(\epsilon^2)}\right] \right)  + \epsilon ~\mathcal{G}_n(\Delta \rho_{n-1}) \nonumber \\
&  = \mathcal{G}_n \left(\frac{e^{-\beta H_n}}{Z_n} \left[ 1 + \epsilon \beta (\Delta H_n - \tr[\Delta H_n]) + O(\epsilon^2)\right] \right)  + \epsilon ~\mathcal{G}_n(\Delta \rho_{n-1}) \nonumber \\
& = \frac{e^{-\beta H_n}}{Z_n} + \epsilon \left[ \mathcal{G}_n(\Delta \rho_{n-1}) + \beta  \mathcal{G}_n(\Delta H_n \frac{e^{-\beta H_n}}{Z_n}) - \beta \tr[\Delta H_n] \frac{e^{-\beta H_n}}{Z_n} \right] + O(\epsilon^2),
\end{align}
\end{widetext}
which reduces to Eq. \eqref{eq:quasistatic} on identifying
\begin{align}
 \Delta \rho_n \equiv& ~ \mathcal{G}_n(\Delta \rho_{n-1}) + \beta  \mathcal{G}_n(\Delta H_n \frac{e^{-\beta H_n}}{Z_n}) \nonumber \\ 
 & - \beta \tr[\Delta H_n] \frac{e^{-\beta H_n}}{Z_n}.
\end{align}

The second part of the proof follows by first obtaining the traceless matrix $\sigma_n$:
\begin{align}
 \sigma_n \equiv & ~\rho_n - \rho_{n-1} = \epsilon \big[ \mathcal{E}_n(\Delta \rho_{n-1}) - \Delta \rho_{n-1} \nonumber 
 \\ & + \beta \mathcal{E}_n(\Delta H_n \frac{e^{-\beta H_n}}{Z_n}) - \Delta H_n  \frac{e^{-\beta H_n}}{Z_n} \big].
\end{align}
Then, we obtain the eigenvalues and eigenvectors of $\rho_n$, given the set $\{ p_n^k, \ket{\psi_n^k} \}$, in terms of the corresponding set for $\rho_{n-1}$, by using the relation $\rho_n = \rho_{n-1} + \epsilon~\sigma_n$. It comes out that
\begin{align} \label{eq:expand1}
p_n^k &= p_{n-1}^k + \epsilon \langle \psi_{n-1}^k | \Delta \sigma_n | \psi_{n-1}^k \rangle + O(\epsilon^2), \\ \label{eq:expand12}
\ket{\psi_n}^k &= \ket{\psi_{n-1}^k} + \epsilon \sum_{l \neq k} \frac{\langle \psi_{n-1}^k | \Delta \sigma_n | \psi_{n-1}^k \rangle}{ p_{n-1}^k - p_{n-1}^l} \ket{\psi_{n-1}^l} \nonumber \\ 
& ~~~ + O(\epsilon^2). 
\end{align}
On the other hand, we may obtain the same quantities to first order in $\epsilon$ from the equation $\rho_n = \frac{e^{-\beta H_n}}{Z_n} + \epsilon \Delta \rho_n + O(\epsilon^2)$. This leads to:
\begin{align}\label{eq:expand2}
p_n^k &= \frac{e^{-\beta E_n^k}}{Z_n} + \epsilon \langle E_n^k | \Delta \rho_n | E_{n}^k \rangle + O(\epsilon^2), \\ \label{eq:expand22}
\ket{\psi_n}^k &= \ket{E_n^k} + \epsilon \sum_{l \neq k} \frac{\langle E_{n}^k | \Delta \rho_n | E_{n}^k \rangle}{ \frac{e^{-\beta E_n^k}}{Z_n} - \frac{e^{-\beta E_n^l}}{Z_n}} \ket{E_n^l} \nonumber \\ 
& ~~~ + O(\epsilon^2),
\end{align}
where the set $\{ E_n^k, \ket{E_n^k} \}$ contains the eigenstates and eigenvectors of the instantaneous Hamiltonian $H_n$.

We are now ready to calculate the change in entropy during the $n$-th step of the process, described by the map $\mathcal{E}_n$. Using Eqs. \eqref{eq:expand1} and \eqref{eq:expand12}, this reduces to:
\begin{align}
& \Delta S_n = S(\rho_n) - S(\rho_{n-1}) = - \sum_k p_{n}^k \ln p_n^k \nonumber \\
& + \sum_k p_{n-1}^k \ln p_{n-1}^k = - \epsilon \sum_k \langle \psi_{n-1}^k | \Delta \sigma_n | \psi_{n-1}^k \rangle \ln p_{n-1}^k \nonumber \\ 
& + O(\epsilon^2).
\end{align}
Finally, by combining Eqs.\eqref{eq:expand1} and \eqref{eq:expand2}, we notice that $p_{n-1}^k = \frac{e^{-\beta E_{n}^k}}{Z_n} + O(\epsilon)$, and hence
$\ln(p_{n-1}^k) = \ln(\frac{e^{-\beta E_n^k}}{Z_n}) + \ln[1 + O(\epsilon)] = \ln(\frac{e^{-\beta E_n^k}}{Z_n}) + O(\epsilon)$. This leads us to write
\begin{align}
 \Delta S_n & = - \epsilon \sum_k \langle \psi_{n-1}^k | \Delta \sigma_n | \psi_{n-1}^k \rangle \ln(\frac{e^{-\beta E_n^k}}{Z_n}) + O(\epsilon^2) \nonumber \\
 &= \epsilon \beta \sum_k E_n^k \langle \psi_{n-1}^k | \Delta \sigma_n | \psi_{n-1}^k \rangle \nonumber \\
 &= - \epsilon \tr[H_n \sigma_n] + O(\epsilon^2),
\end{align}
where, in the last step, we have used $\ket{p_{n-1}^k} = \ket{E_n^k} + O(\epsilon)$ following from Eqs. \eqref{eq:expand12}
and \eqref{eq:expand22}. The relation above corresponds to Eq. \eqref{eq:reversible}, thus completing the proof. 

\subsection{Arbitrary initial states}

Here we consider in detail the case in which $\rho_0$ is not full rank, that is, it contains one or more eigenstates with associated eigenvalues equal to zero. 
In such case, since $\rho_0$ is not invertible, $\ln \rho_0$ is not well defined, and a modification of the definition of $H_0$ is needed. In such case we set
\begin{equation} \label{eq:def}
 H_0 \equiv - k_B T \ln \left( \rho_0 + \alpha \mathbb{I}_\mathrm{ker} \right),
\end{equation}
where $\alpha \ll 1$ is a (small) real number, and $\mathbb{I}_\mathrm{ker}$ is the identity on the kernel (or null space) of $\rho_0$. If the state $\rho_0$ is full rank 
then $\mathbb{I}_\mathrm{ker} = 0$ and we recover the previous expression. The definition \eqref{eq:def} implies that now the initial arbitrary state $\rho_0$ can be rewritten as 
\begin{equation}\label{eq:ini}
 \rho_0 = \frac{e^{-\beta H_0}}{Z_0} + \alpha \Delta \rho_0,
\end{equation}
where $Z_0 = 1 + \alpha d_\mathrm{ker}$, with $d_\mathrm{ker}$ the dimension of the kernel (the nullity of $\rho_0$), and 
$\Delta \rho_0 = \exp(- \beta H_0)\frac{d_\mathrm{ker}}{Z_0} - \mathbb{I}_\mathrm{ker}$ a traceless operator.

This means that the only difference with respect to the previous situation is that now after the initial quench of the work extraction protocol, that is, at the beginning of the quasi-static process, 
the state of the system is not exaclty in equilibrium with the reservoir, but $\alpha$-close to it in the sense of Eq. \eqref{eq:ini}. However, as we have seen before, our protocol is robust to small departures from equilibrium. 
More precisely, if $\alpha$ is of the order of $\epsilon$ ---the order of the small changes in the Hamiltonian during each infinitesimal step of the protocol--- the initial state fulfills Eq. \eqref{eq:quasistatic}, 
for which the reversibility condition in Eq.\eqref{eq:result} follows. 

In practice, Eq. \eqref{eq:def} means that one needs to lift the eigenvalues of $H_0$ corresponding to the kernel of $\rho$, let us call them the set $\{ E^{\mathrm{ker}}_n\}_{n=1}^{d_\mathrm{ker}}$, 
so that the corresponding thermal occupations $\exp(-\beta E_n^{\mathrm{ker}})/Z_0$ are negligible, that is $E_n^{\mathrm{ker}} \gg k_B T$, which is equivalent to $\alpha \ll 1$. Notice that this does not cost any work since the occupation probabilities 
of those eigenstates in $\rho_0$ are strictly zero. We also notice that our proposal is an extension to the quantum realm of what is done in a more classical context. For example, in some experimental realizations of the Szilard engine, an equivalent protocol is achieved by 
modulating the profile of a potential trapping a single particle \cite{Pekola,Roldan}. In those configurations a double well potential is used to encode two informational states. If the particle is detected in one of the wells, then the potential rapidly changes adapting 
to an effective single well potential trapping the particle in his actual state, which becomes the equilibrium distribution \cite{Par15}. 

\section{Isothermal reversible work extraction outperforms ergotropy for any temperature}

In this appendix we prove that the work that can be extracted from a system with Hamiltonian $H$ in an arbitrary nonequilibrium state $\rho$  by the optimal reversible isothermal cycle (in which access to a thermal reservoir at any temperature is allowed) is always greater than that obtained when only unitary operations can be performed (given by ergotropy). In the first case, the maximum extractable work
is given by the nonequilibrium free energy difference between the initial state $\rho$ and the equilibrium state at the reservoir temperature, $\rho^\beta = e^{-\beta H}/Z$,
where $\beta = 1/k_{B} T$ is the inverse temperature and $Z = \tr[e^{-\beta H}]$ the partition function \cite{Par15}. That is
\begin{equation}
 {\cal W}^\beta = \mathcal{F}_\beta (\rho) - \mathcal{F}_\beta (\rho^\beta),
\end{equation}
where $\mathcal{F}_\beta (\sigma) \equiv \tr[H \sigma] - k_B T S(\rho)$ is the definition of the nonequilibrium free energy with respect to $\beta$, $S(\sigma)$ denoting the von Neumann
entropy.

On the other hand, when only unitary operations are available, the maximum extractable work is given by the ergotropy $\mathcal{W}$, which is always bounded by \cite{All04}
\begin{equation} \label{eq:ergo}
\mathcal{W} \leq \tr[H \rho] - \tr[H \rho^{\beta_\ast}] \equiv \mathcal{W}_\mathrm{max},
\end{equation}
where $\beta_\ast$ is that particular temperature which guarantees that entropy is unchanged: $S(\rho) = S(\rho^{\beta_\ast})$. We notice that $\beta_\ast$ is obtained when
minimizing the energy of the system for a fixed entropy.

In the following, we will prove that
\begin{equation} \label{eq:theorem}
 \mathcal{W}^\beta \geq \mathcal{W}_\mathrm{max} ~~~ \forall~ \beta \in \mathbb{R}^{+},
\end{equation}
and that the equality is reached if and only if $\beta = \beta_\ast$.

We start by analyzing the quantity $\mathcal{W}^\beta$, to notice that
\begin{align} \label{eq:1}
\mathcal{W}^\beta &=  \tr[H \rho] - \tr[H \rho^\beta] - k_B T [S(\rho) - S(\rho^\beta)] \nonumber \\
         &= \mathcal{W}_\mathrm{max} + \tr[H \rho^{\beta_\ast}] - \tr[H \rho^\beta] \nonumber \\ 
         & ~~~ - k_B T [S(\rho) - S(\rho^\beta)],
\end{align}
where we simply added and subtracted $\tr[H \rho^{\beta_\ast}]$ and identified the maximum ergotropy from Eq. \eqref{eq:ergo}. Now, we notice that, by definition,
$S(\rho) = S(\rho^{\beta_\ast})$, so that Eq. \eqref{eq:1} can be rewritten as
\begin{equation} \label{eq:2}
 \mathcal{W}^\beta =  \mathcal{W}_\mathrm{max} + \mathcal{F}_\beta (\rho^{\beta_\ast}) - \mathcal{F}_\beta (\rho^\beta).
\end{equation}
Finally, by noticing that $\mathcal{F}_\beta (\rho) - \mathcal{F}_\beta (\rho^\beta) = k_B T D(\rho || \rho^\beta)$ for any initial state $\rho$, we immediately
obtain:
\begin{equation} \label{eq:relation}
 \mathcal{W}^\beta = \mathcal{W}_\mathrm{max} + k_B T D(\rho^{\beta_\ast} || \rho^\beta ) \geq \mathcal{W}_\mathrm{max},
\end{equation}
where $D(\rho || \sigma) = \tr[\rho (\ln \rho - \ln \sigma)]$ is the quantum relative entropy. Eq. \eqref{eq:theorem} then follows from the non-negativity of the quantum relative entropy,
$D(\rho || \sigma) \geq 0$, and the fact that the latter is  zero only for $\rho = \sigma$ (Klein's inequality \cite{Nielsen}).
Consequently, we have that $\mathcal{W}^\beta = \mathcal{W}_\mathrm{max}$ only when $\beta = \beta_\ast$, which completes the proof.

\section{Reversible work extraction from quantum correlations}

In the main text we propose a protocol for extracting work from quantum correlations based on the inclusion of a new initial step before projective measurements on the ancilla are performed.
Here, we provide some intuitive considerations motivating the construction of this initial sub-process. First, it should be noticed that we are looking for a reversible process leaving the total system in a state with zero quantum correlations, but with the same classical correlations as $\rho_{SA}$. As mentioned in the main text, this means requiring the final state after this initial sub-process
to be
\begin{equation}
 \rho_{SA}^\prime = \sum_k p_k \rho_{S|\Pi_A^k} \otimes \Pi_A^k,
\end{equation}
with $\Pi_A^k$ belonging to the optimal set of projectors.

In order to construct a reversible stroke between $\rho_{SA}$ and $\rho_{SA}^\prime$, we may first imagine the two following steps:
i) apply to the whole $SA$ system an isothermal protocol that allows for the reversible conversion of $\rho_{SA}$ into the thermal product state $\rho_{S}^{\beta} \otimes \rho_{A}^{\beta}$
[henceforth extracting an amount of work ${\mathcal W}^{\beta}_{SA}$ as given in Eq. (9) of the main text]. This includes the sudden
quench of the total Hamiltonian $H_S + H_A \rightarrow H_{SA} \equiv -k_B T \ln \rho_{SA}$ and a quasi-static driving of the full system as explained in the first section of this Supplemental Material;
ii) perform the reversible transformation of $\rho_{S}^{\beta} \otimes \rho_{A}^{\beta}$ into $\rho_{SA}^\prime$ by simply inverting the isothermal protocol which should be applied for
optimal work extraction from state $\rho_{SA}^\prime$ and leading to $\rho_{S}^{\beta} \otimes \rho_{A}^{\beta}$. This means that we need to apply the quasi-static driving
$H_{S} + H_{A} \rightarrow H_{SA}^\prime \equiv - k_B T \ln \rho_{SA}^\prime$, followed by a final quench $ H_{SA}^\prime \rightarrow H_{S} + H_{A}$.
Notice that step ii) requires {\it performing} the work
\begin{eqnarray}
W_\mathrm{in} & = &
\mathcal{F}_{\beta}(\rho_{SA}^\prime)- \mathcal{F}_{\beta}(\rho_{S}^{\beta} \otimes \rho_{A}^{\beta})
= \nonumber \\
& = & {\mathcal W}^{\beta}_S + {\mathcal W}^{\beta}_A + k_B T \mathcal{J}(\rho_{SA}).
\end{eqnarray}

From the above considerations, we find that the reversible stroke we are searching for can be obtained e.g. by combining steps i) and ii), since both of them are reversible and their combination leaves the system in
$\rho_{SA}^\prime$. Moreover, we notice that they can be merged since i) ends with a quasi-static process leading to $\rho_{S}^{\beta} \otimes \rho_{A}^{\beta}$ while ii) starts with a
quasi-static process from the same state. As a result, we obtain the process introduced in the main text: A sudden quench of the total Hamiltonian, $H_S + H_A \rightarrow H_{SA} \equiv -k_B T \ln \rho_{SA}$,
followed by a quasi-static driving transforming $H_{SA} \rightarrow H_{SA}^\prime$, which leads the compound system to end up in the state $\rho_{SA}^\prime$, as it follows
from the fact that $\rho_{SA}^\prime$ is now the equilibrium state at temperature $T$ with respect to the Hamiltonian $H_{SA}^\prime$. Finally, a second
sudden change of the Hamiltonian is performed $H_{SA}^\prime \rightarrow H_{S} + H_{A}$. We notice that, in order to achieve each of these three steps, one would need not only to control system
and ancilla locally, but also to manipulate their interaction in a suitable way.


\begin{thebibliography}{99}
\bibitem{noteref}
This implies that, if the driving protocol is stopped at some arbitrary intermediate time, the system would thermalize to the Gibbs state defined by its actual Hamiltonian. At the same time, this implies that the system-reservoir interaction may need to change in form in order to give rise to such a thermalization process.
\bibitem{elouard}
Recently, it has been shown that this quantity could be negative for some Maxwell's demon engines, see Ref. \cite{extradem}.
\end{thebibliography}
\end{document}